\documentclass[tighten,twocolumn,apj,onecolumnappendix]{likeapj}
\usepackage{float}
\usepackage{multirow}

\usepackage{apjfonts}
\shorttitle{SMG-QSO \Lya\ Nebula}
\shortauthors{Hall \& Fu}
\usepackage{lineno}

\newcommand{\kms}{{km\,s$^{-1}$}}
\newcommand{\ergps}{{erg\,s$^{-1}$}}
\newcommand{\msun}{$M_\odot$}
\newcommand{\msunyr}{$M_\odot\,{\rm yr}^{-1}$}

\newcommand{\um}{$\mu$m}

\newcommand{\DC}{{\it CubeCarve}}

\newcommand{\surf}{erg\,s$^{-1}$\,cm$^{-2}$\,arcsec$^{-2}$ }

\newcommand{\Lya}{Ly$\alpha$}

\newcommand{\HI}{H\,{\sc i}}

\newcommand{\OII}{[O\,{\sc ii}]}

\newcommand{\idOne}{G15 1419$+$0052} 
\newcommand{\idTwo}{G15 1450$+$0026} 
\newcommand{\idThree}{G09 0902$+$0101} 
\newcommand{\idFour}{G15 1444$-$0044} 

\newcommand{\SQSO}{SMG-QSO}

\newcommand{\mycomment}[1]{}
\usepackage{amsmath}
\usepackage{amssymb}
\defcitealias{Fu17}{F17}
\defcitealias{Gonzalez-Lobos23}{G23}
\defcitealias{Hall24}{H24}

\begin{document}

\title{Mapping the Extended Lyman-Alpha Emission within the Circumgalactic Medium of Quasars Hosted by Dusty Starbursts with {\it CubeCarve} }
\author{
Kevin~Hall and Hai~Fu
}
\affiliation{Department of Physics \& Astronomy, University of Iowa, Iowa City, IA 52242}

\begin{abstract}

{We present a study of extended Ly$\alpha$ emission around four quasars hosted by dusty starbursts, which are composite systems thought to represent a transitional stage in quasar evolution. To extract faint CGM emission in the presence of bright point sources, we introduce {\it CubeCarve}, a dual-channel deconvolution algorithm that separates unresolved quasar emission from spatially extended structure. This approach enables reliable recovery of \Lya\ emission projected onto the quasar position without introducing subtraction artifacts. Using {\it CubeCarve}, we find that the \Lya\ surface brightness profiles of these systems are, on average, fainter and shallower than those of quasars of similar bolometric luminosities. We also find that the total integrated \Lya\ luminosities of the nebulae are lower in systems whose host galaxies exhibit brighter far-infrared emission. These results suggest that the CGM conditions in composite systems differ from those in the broader quasar population. Our study highlights both the physical diversity of quasar CGM environments and the effectiveness of {\it CubeCarve} for recovering diffuse emission in modern IFU datasets.}

\end{abstract}


\section{Introduction}

The circumgalactic medium (CGM) is the region between a galaxy’s interstellar medium (ISM) and the surrounding intergalactic medium (IGM) \citep{Tumlinson17}. It serves as a crucial laboratory for studying the processes that govern galaxy growth and evolution. In the standard cosmological picture, dark matter halos assemble over time through continuous accretion of gas, leading to the formation of galaxies within them. In the most massive halos ($M_{\rm DM}\sim 10^{12.5}$\,\msun), the accreting gas encounters shocks and is heated to virial temperature ($T_{\rm vir}\sim 10^6$\,K) \citep[e.g.,][]{Rees77, Silk77, White78, Birnboim03}. Yet, despite these high temperatures, galaxies embedded in such halos continue to grow and sustain active star formation. This persistence requires the presence of a cooler ($\sim 10^4$\,K) gas component to replenish the interstellar medium, as the cooling time of the X-ray emitting hot gas is far too long to support the observed star formation rates (SFRs). The CGM thus provides a unique window into detecting this cool gas and understanding the mechanisms that regulate the evolution of massive galaxies.

Submillimeter-bright galaxies \citep[SMGs;][]{Smail97,Barger98,Blain02} are among the most extreme star-forming systems in the Universe, as they typically have star formation rates exceeding $500$\,\msunyr. Their extreme infrared luminosities, observed between 850\,\um\ and 2\,mm, arise from dust that reprocesses the intense radiation from young, massive stars. This emission is further amplified by the negative $K$-correction in the Rayleigh$-$Jeans tail, making SMGs readily detectable even at $z\sim2$--3 \citep{Chapman05}. Clustering analyses suggest that SMGs inhabit massive dark matter halos ($M_{\rm DM}\sim9\times10^{12}$\,\msun; \citealt{Hickox12}), indicating that they are precursors to today’s massive ellipticals. The halos surrounding SMGs are expected to host both hot virialized gas and possibly cool, dense filaments of accreting material from the cosmic web \citep{Keres05,Keres09,Dekel06,Dekel09,Faucher-Giguere11,Martin19}. Studying the CGM of SMGs thus offers insight into how massive galaxies acquire and process gas during their peak star-forming phases.

While SMGs represent a key phase of rapid stellar growth, quasi-stellar objects (quasars/QSOs) offer an alternative and complementary view of the CGM by serving as bright background beacons. As active galactic nuclei (AGN), their luminous continuum emission serve as backlights against which intervening gas imprints characteristic absorption features. QSOs have been utilized to probe the CGM of other QSOs. It is through these observations where their CGMs have been found to possess a large supply of cool, metal-entriched gas \citep{Hennawi06,Prochaska14}. Indeed, QSOs are also found to reside in similarly massive DM halos ($M_{\rm DM} \sim 10^{12.5}$\,\msun) \citep{White12} as SMGs. This suggests there may be an evolutionary link between SMGs and QSOs, so studying both of their CGMs is of great interest. However, identifying projected SMG/QSO pairs suitable for such studies remains observationally challenging.

Early sub-millimeter observations revealed that QSOs are among the brightest sources at these wavelengths, rivaling SMGs \citep{Isaak94,McMahon94}. Single-dish facilities, such as the Institut de Radioastronomie Millimétrique (IRAM) 30\,m telescope and the James Clerk Maxwell Telescope (JCMT), provided early detections but with limited spatial resolution (FWHM $\approx10\farcs6$ at 1.2\,mm for IRAM and $13\farcs8$ at 850\,\um\ for JCMT). These beam sizes hindered the identification of the true far-IR counterparts to optically selected QSOs. Interferometric observations have since resolved these ambiguities, revealing several categories of systems: (1) QSOs hosted by dusty starbursts  \citep[\SQSO\ composite galaxies][]{Trakhtenbrot17,Fu17}; (2) gas-rich mergers involving interacting \SQSO\ systems \citep{Omont96,Carilli02,Clements09,Trakhtenbrot17}; (3) “wet-dry” mergers between gas-poor QSOs and gas-rich SMGs; and (4) projected \SQSO\ pairs. Projected pairs, such as the GAMA\,J0913$-$0107\,system \citep{Fu21}, enable rare background-probing of the CGM. However, absorption-based studies probe only isolated sightlines. To get a better view of the CGM, we must directly map its emission.

Direct searches for diffuse emission have focused on bright rest-UV lines such as H\,{\sc i} Lyman-$\alpha$ (\Lya) emission line, which traces the cool gas. Theoretical models predicted that radiation from star formation and accretion onto central black holes could generate a “\Lya\ glow” within the CGM \citep{Rees88,Haiman01}. Narrowband imaging surveys targeting overdense environments revealed luminous “\Lya\ blobs” (LABs) with physical extents of hundreds of kpcs and total luminosities $L_{\rm Ly\alpha} > 10^{43}$ erg\,s$^{-1}$ \citep[e.g.,][]{Steidel00,Matsuda04,Dey05,Nilsson06}. Early attempts to explain this emission invoked gravitational cooling radiation from infalling gas \citep{Goerdt10,Rosdahl12}, though subsequent studies suggested that intense star formation or AGN activity within these regions can also account for the observed luminosities \citep{Chapman01,Geach14,Geach16,Umehata21}.

QSOs provide a particularly effective means of illuminating the CGM, as their intense ionizing radiation can generate copious amounts of \Lya\ photons through recombination in surrounding gas \citep{Cantalupo05,Cantalupo12,Kollmeier10,Hennawi13}. In addition, \Lya\ photons emitted from the broad-line region (BLR) can resonantly scatter off neutral hydrogen, further enhancing the diffuse glow \citep{Cen13,Hennawi13,Gronke17,Byrohl22}. Narrowband imaging revealed numerous giant \Lya\ nebulae with surface brightnesses exceeding $10^{-17}$ erg s$^{-1}$ cm$^{-2}$ arcsec$^{-2}$ \citep[e.g.,][]{Heckman91a,Cantalupo14,Hennawi15,Cai17}, but fixed filter passbands and velocity offsets between the QSO and nebular emission often limited the recovered flux and structure. These challenges motivated the development and use of integral-field spectroscopy towards mapping the CGM.

The advent of high-throughput integral-field units (IFUs) such as the Keck Cosmic Web Imager \citep[KCWI;][]{Morrissey18} and the Multi Unit Spectroscopic Explorer \citep[MUSE;][]{Bacon10} has enhanced our view of \Lya\ emission in QSO environments. These instruments enable mapping of faint, diffuse emission down to surface brightness levels of $\sim10^{-18}$\,erg\,s$^{-1}$\,cm$^{-2}$\,arcsec$^{-2}$. Surveys using MUSE and KCWI have revealed hundreds of \Lya\ nebulae with diverse morphologies and kinematics around QSOs spanning $2<z<6$ \citep{Borisova16,Arrigoni-Battaia18,Cai18,Arrigoni-Battaia19,Cai19,Farina19,Lau22,Vayner23,Gonzalez-Lobos23}. Nearly all bright QSOs show extended \Lya\ halos, with gas motions consistent with gravitationally bound material. Comparing $z>3$ MUSE samples to $z\sim2$ KCWI observations reveals fainter average surface brightness at lower redshift, likely reflecting a decline in the cool-gas covering factor \citep{Cai19}. Despite these insights, the physical link between the QSO and its surrounding \Lya\ emission remained only partially understood.

Building on these results, the third installment of the QSO MUSEUM survey \citep{Arrigoni-Battaia19,Herwig24,Gonzalez-Lobos25} explored how \Lya\ nebula properties scale with QSO luminosity, targeting QSOs with bolometric luminosities spanning $45.1 < \log(L_{\rm bol}/[\rm erg,s^{-1}]) < 48.7$. This effort built on earlier work by \citet{Mackenzie20}, who found that faint QSOs host less luminous \Lya\ nebulae, suggesting a luminosity dependence. By substantially expanding the faint end of the luminosity distribution, \cite{Gonzalez-Lobos25} demonstrated that brighter QSOs generally host \Lya\ nebulae with higher surface brightness, highlighting the important role of ionizing photons from the active nucleus. Together, these findings point to a direct link between AGN radiative output and the surrounding CGM.

If we pivot from QSO CGM studies and instead consider SMGs, we gain an opportunity to isolate the physical mechanisms that drive extended \Lya\ emission. SMGs generate abundant ionizing photons through intense, obscured star formation, but their large dust reservoirs may severely attenuate \Lya\ emission. Indeed, two $z>4$ SMGs observed with MUSE \citep{Gonzalez-Lobos23} show no detectable \Lya\ emission above the achieved surface brightness limits. In contrast, SMGs residing in protocluster environments, where the ambient UV radiation field is likely elevated, have been observed to host extended \Lya\ halos \citep{Umehata19}. These mixed results suggest that ultra-deep observations may be required to reveal \Lya\ emission around isolated SMGs. For example, deep KCWI observations ($t_{\rm exp} > 2$\,hr) of the GAMA\,J0913$-$0107 system \citep{Hall24} uncovered a filamentary \Lya\ structure consistent with a cool, infalling stream embedded within a massive hot halo.

A particularly powerful regime for CGM studies arises when targeting SMGs that are transitioning into QSOs, the so-called SMG-QSO composite systems. These objects offer a unique laboratory in which both star formation and black hole accretion contribute to ionization, feedback, and metal enrichment, potentially imprinting distinct signatures on their \Lya\ nebulae. This line of investigation began with MUSE observations of two SMG-QSOs at $z\sim3$ \citep{Gonzalez-Lobos23}, which revealed relatively faint integrated \Lya\ nebula luminosities compared to a more luminous QSO. In this work, we extend this investigation using KCWI observations of two additional SMG-QSOs at $z\sim2.7$, enabling a systematic comparison of the extended \Lya\ properties across the four known systems.

To support this effort, we address a key methodological challenge in IFU studies of QSOs. Strong subtraction residuals are commonly found within the inner $\lesssim 20$\,kpc of the nebula when removing the bright QSO. While this limitation is negligible for extremely bright and extended \Lya\ halos, it becomes critical when studying fainter systems. To overcome these limitations, we introduce {\it CubeCarve}, a new dual-channel deconvolution method that separates unresolved QSO emission from extended nebula emission. This enables us to robustly recover \Lya\ morphology and surface brightness profiles at small radii, providing an artifact-free view of the CGM.

The paper is organized as follows. Section\,\ref{sec:targets} describes the SMG-QSO sample and the observational setup. In Section\,\ref{sec:algorithm}, we introduce the algorithm underlying {\it CubeCarve}, validate its performance, and describe its application to our observations. In Section\,\ref{sec:results}, we present the properties of the \Lya\ nebulae associated with each target. We then quantify how SMG-QSOs differ from the broader QSO population by constructing \Lya\ surface brightness profiles and examine how host galaxy properties may be connected to \Lya\ emission in Section\,\ref{sec:comparisons}. We conclude with final remarks in Section\,\ref{sec:conclusions}.

Throughout, we adopt a flat $\Lambda$CDM cosmology with $\Omega_{\rm m} = 0.3$, $\Omega_{\Lambda}=0.7$, and $h \equiv H_0 / (100\,{\rm km\,s^{-1}\,Mpc^{-1}})$.

\section{SMG-QSOs}\label{sec:targets}

This work continues our KCWI investigation of extended \Lya\ emission within the CGM of high-redshift galaxies. In our previous paper, we presented observations of the GAMA\,J0913$-$0107 system in \cite{Hall24} (hereafter \citetalias{Hall24}), a rare projected alignment between a SMG and two background QSOs. The overarching goal of this program is to obtain deep, spatially resolved observations of the CGM surrounding SMGs, thereby probing the physical mechanisms that regulate galaxy growth and evolution. The work presented here represents the culmination of our KCWI campaign, extending the search for extended emission to a broader class of dusty, star-forming galaxies hosting luminous QSOs.

\subsection{Target Selection}

\begin{figure*}[!htb]
    \centering
    \includegraphics[width=\textwidth]{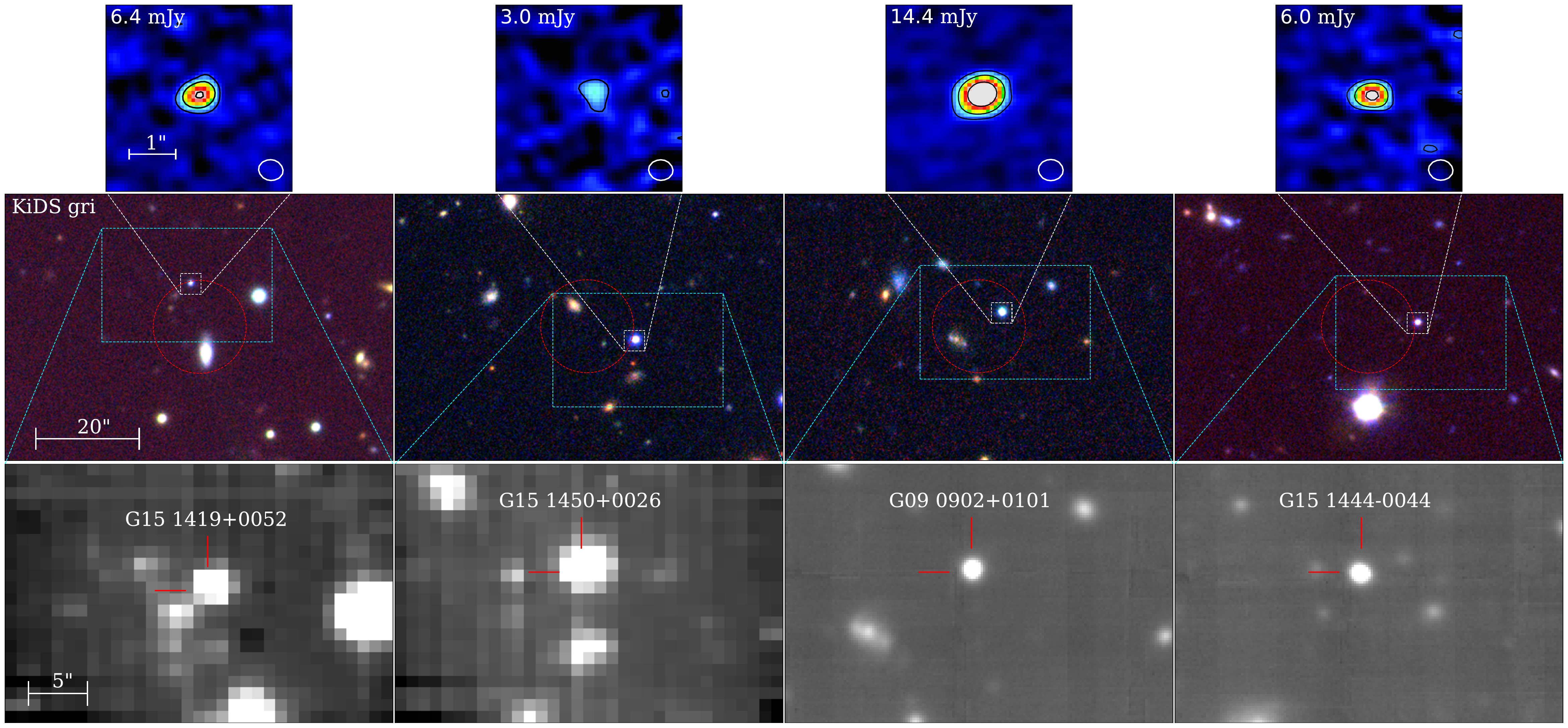}
    \caption{The finder chart for each \SQSO\ composite system. {\it Middle Panel}: The KiDS gri tri-color image of the field centered at the position of the Herschel $250$\,\um\ source. The red-dashed circle corresponds to its positional uncertainty ($12\farcs5$). {\it Top Panel}: We provide the ALMA $870$\,\um\ continuum maps of each Herschel-detected source. Each panel is $4\farcs \times 4\farcs$, and we indicate its location relative to the KiDS image as a white box. Additionally, the ellipse in the lower-right corner represents the shape of the CLEAN beam. {\it Bottom Panel}: We present the pseudo-$g$-band images of the field as seen through KCWI and MUSE.    
    }
    \label{fig:finder-chart}
\end{figure*}

\cite{Fu16} cross-matched a large sample of spectroscopically confirmed QSOs with {\it Herschel} sources detected in SPIRE bands (250, 350 and 500\,$\mu$m). In particular, they selected sources peaking at $350$\,$\mu$m, the so called “350\,$\mu$m peakers”, with an accompanying $S_{500} > 20$\,mJy. Using the Rayleigh$-$Jeans extrapolation, these correspond to submillimeter sources with $S_{870} \gtrsim 3$\,mJy.

To improve the {\it Herschel} astrometry and facilitate spectroscopic follow-up, \cite{Fu17} (hereafter \citetalias{Fu17}) obtained ALMA Band-7 (345\,GHz/870\,$\mu$m) imaging of 29 SMG-QSO pairs with angular separations $5\arcsec < \theta < 30\arcsec$. The superior resolution (FWHM $\sim 0\farcs5$) revealed four bright 870\,$\mu$m sources ($S_{870} \gtrsim 3$\,mJy) coincident with the QSO optical positions. These are \idOne\ ($z=2.6711$), \idTwo\ ($z=2.8220$), \idThree\ ($z=3.1204$), and \idFour\ ($z=3.3750$), which define the sample of \SQSO\ composites studied here. Finder charts for these systems are shown in Figure \ref{fig:finder-chart}, including the ALMA continuum maps that emphasize the positional overlap.

Host galaxy properties were derived by \citetalias{Fu17} through SED fitting, combining SDSS photometry ($3500-9000$\,\AA), {\it WISE} mid-IR data ($3.4-22$\,$\mu$m), {\it Herschel}/SPIRE photometry, and ALMA 870\,$\mu$m imaging. A standard QSO template reproduced the SDSS and {\it WISE} photometry well. Despite the strong submillimeter emission from their dusty hosts, the best-fit QSO SEDs show no evidence of dust reddening (see Figure\,5 in \citetalias{Fu17}). This raises the possibility that these QSOs were once heavily obscured and are now caught in a transitional phase, evolving from a dust-enshrouded starburst toward a quasar-dominated system. A natural question then is whether their CGM differs from that of “standard” QSOs \citep[e.g.,][]{Arrigoni-Battaia19,Cai19}. To address this, we obtained deep KCWI observations of the $z < 3$ \SQSO\ targets and reanalyzed archival VLT/MUSE observations for the $z > 3$ systems.

\subsection{Observations \& Data Reduction}

\begin{figure}[!htb]
\hspace*{-0.4cm}
	\includegraphics[width=0.5\textwidth]{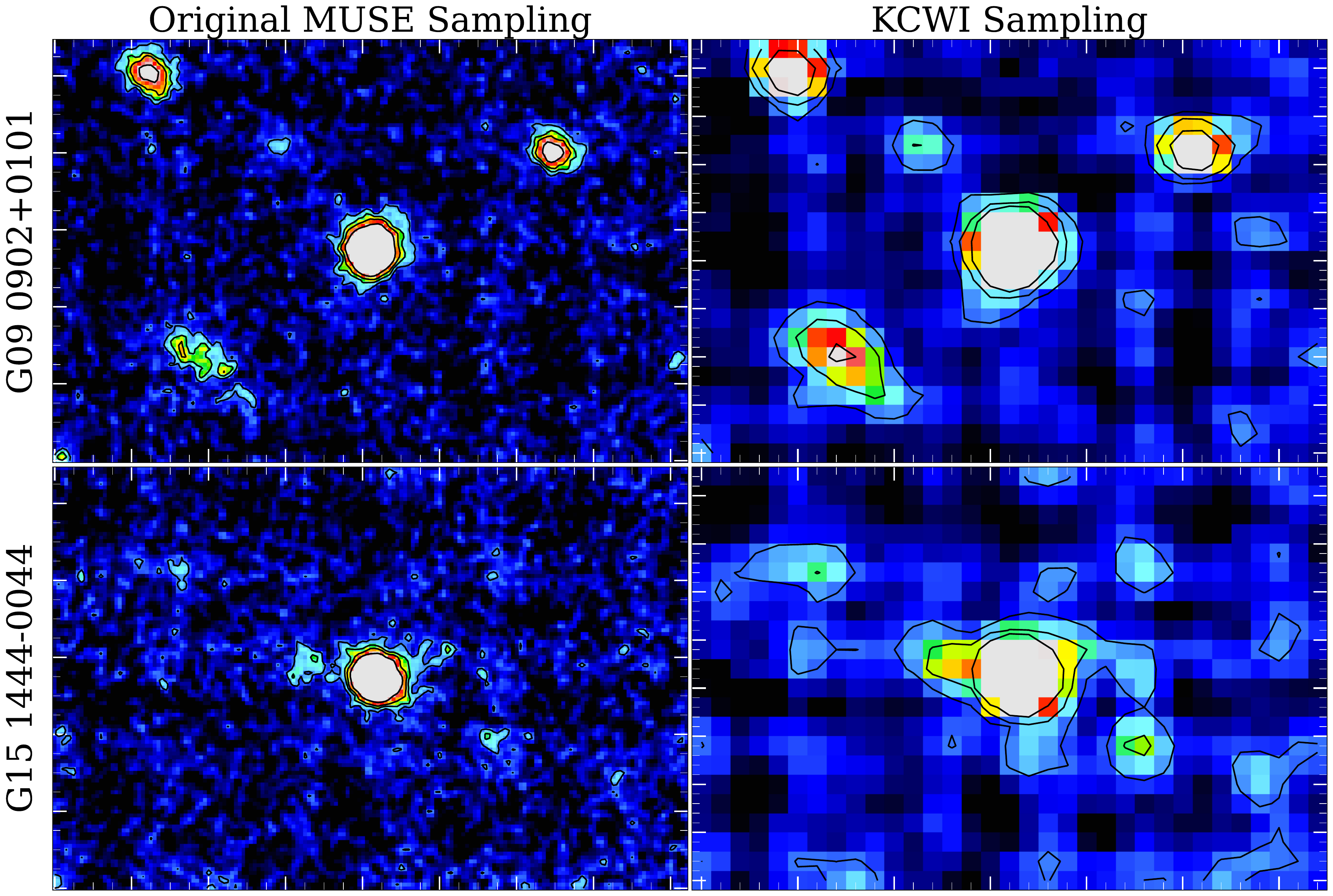}
	\caption{A visual demonstration of the MUSE resampling procedure to generate ``KCWI-like'' datacubes. Each panel is an integrated surface brightness map across a $\delta v = \pm 1000$\,\kms. {\it Left}: Original MUSE pixel sampling. {\it Right}: Resampled  MUSE datacube.}
	\label{fig:resampling}
\end{figure}

We carried out deep optical integral-field spectroscopy with Keck II/KCWI on Jan 30, 2022 (UT). A single configuration was used throughout the night. Specifically, we utilized the large image slicer along with the blue medium (BM) grating tilted to a central wavelength of 4500\,\AA, and with a 2x2 binning on the CCD detector. We achieved a wavelength coverage of $4055$\,\AA\ $ < \lambda < 4940$\,\AA\, with a dispersion of $\sim$0.5\,\AA/pix and a spectral resolution of R $\sim 2000$. The wavelength coverage is nearly $g-$band and includes the \HI\ Ly$\alpha$ line between $2.33 < z < 3.06$, capturing the redshifts of both \SQSO\ composite systems at $z < 3$. Additionally, our coverage enables the detection of the \OII\ $\lambda\lambda$3727,30 doublet from interloper galaxies between $0.088 < z < 0.325$. 

For both targets, we position the QSO at the center of the 33\farcs0 wide (24$\times$1.35\arcsec\ slits) and 20\farcs4 tall (the length of each slit) field-of-view (FOV) obtained with the large image slicer of KCWI. For each science exposure, we kept a $0^\circ$ (i.e., N is up) position angle (PA). We performed $4 \times 20$ min (1.33 hr) exposures for \idOne, and $5 \times 20$ min (1.66 hr) exposures for \idTwo. Between each exposure, we dithered between the base position and $3\farcs0$ to the South. The atmospheric seeing throughout the night was $\approx 1\farcs2$. 

Data reduction on both targets was carried out with the Python version (Release 1.1) of the KCWI data reduction pipeline\footnote{\url{https://github.com/Keck-DataReductionPipelines/KCWI_DRP}} \citep{neill18}. We opted to use the dome flats for the calibration because it left fewer systematic artifacts on the sky-subtracted 2D spectra. We used Feige\,56 for flux calibration as it has less absorption features near our wavelength range. The pipeline produces fully calibrated, differential atmospheric refraction corrected, datacubes $(x,y,\lambda)$ and variance cubes.

To stack the individual frames, we first solve for the astrometry by matching the positions of the sources detected in the wavelength-collapsed KCWI pseudo-$g$-band image with their equatorial coordinates from the KiDs catalog. We aligned each KCWI frame using the QSOs in each \SQSO\ system. For \idOne, an additional bright foreground star was chosen to improve the astrometric solution. We used a modified version of the Cosmic Web Imager Tools \citep[CWITools; ][]{OSullivan20} to coadd the datacubes based on their astrometric solution. The python package combines the datacubes by re-sampling the pixels onto a common grid using the {\it drizzle} algorithm \citep{Fruchter02}. However, our dither pattern does not enable a significant improvement on the spatial resolution, so we opted for a larger pixel size of $1\farcs0 \times 1\farcs0$ to match the typical seeing; any pixel size smaller led to noticeable artifacts. To illustrate the stacking procedure, we produced a pseudo-$g$-band image from the coadded KCWI datacube by averaging all wavelength channels. These are presented in Figure\,\ref{fig:finder-chart}, and we recover the sources detected in the deep $g$-band image from KiDS that reached a 5$\sigma$ detection limit of 25.4 magnitude, despite its lower resolution. 

To produce the coadded variance cube for S/N estimation, we first pass the variance cubes through the same {\it drizzle} procedure just like the flux cubes. To account for covariances introduced to the coadded datacube from resampling, we use the CWITools covariance estimation function to compute the factor that should be multiplied to the resampled variance cube \citep[refer to Section 5.2 in ][for details]{OSullivan20a}. For our case, we found a multiplication factor of $\sim 2$.

The remaining $z>3$ \SQSO\ targets have their \Lya\ emission lines outside of the spectral coverage of KCWI's blue channel. However, both \idThree\ and \idFour\ were previously observed by MUSE/VLT. For this work, we downloaded the fully reduced and stacked MUSE datacubes from the ESO data portal under the Program ID 0102.A-0403(A). This configuration achieved a wavelength coverage of $4750$\AA\, $< \lambda < 9350$\,\AA\, with R $\approx 2230$ at 6078\,\AA. The data portal performed an automatic data reduction following the MUSE pipeline version 2.4 \citep{Weilbacher20}. Additional sky residuals were subtracted out from each cube using the Zurich Atmospheric Purge (ZAP) software \citep{soto16}. We refer to \cite{Gonzalez-Lobos23} for full details on the observation. Since we are only interested in comparing the extended \Lya\ emission properties for all four \SQSO\ targets, we save on computational resources by reducing the spectral coverage down to $4750$\AA\, $< \lambda < 6624$\AA. Additionally, we decrease the full $1' \times 1'$ MUSE FOV down to 33\farcs0 $\times$ 20\farcs4 to match our KCWI configuration. The pseudo-$g$-band images for both MUSE targets are presented in the bottom panel of Figure \ref{fig:finder-chart}.

To enable a fair comparison between the MUSE and KCWI observations, we must account for instrumental differences that affect the spatial sampling of the datacubes. As described above, we first restrict the MUSE field of view to match that of KCWI. More importantly, the two instruments operate at different spatial resolutions. MUSE has a finer sampling of $0\farcs2 \times 0\farcs2$ per pixel, whereas the final coadded KCWI datacubes have a coarser $1\farcs0 \times 1\farcs0$ scale. To place both datasets on the same footing, we resample the MUSE datacubes to the KCWI pixel grid.

To generate "KCWI-like" datacubes, we first convolve each MUSE datacube with a 2D Gaussian kernel resembling the seeing during the night of our observations. We use a kernel size equal to $3\times$ the width of the FWHM, which corresponds to $15 \times 9$\, MUSE pixels. After convolution, we rebin the MUSE datacube to the KCWI pixel scale. Specifically, one KCWI pixel ($1\farcs0 \times 1\farcs0$) corresponds to $5 \times 5$ MUSE pixels, since $5 \times 0\farcs2 = 1\farcs0$. We therefore sum the flux within each $5 \times 5$ block of MUSE pixels to define the rebinned value at the KCWI scale, repeating this procedure for every spectral channel. Figure\,\ref{fig:resampling} presents surface brightness images of \idThree\ and \idFour\ after the resampling procedure. As one can see, the bright QSOs of \idThree\ and \idFour\ now resemble KCWI observations.

An important detail to note is the impact on the MUSE variance cube. Once the datacubes are convolved and resampled to the KCWI grid, additional covariance is introduced that must be accounted for prior to any extraction of extended \Lya\ emission. We empirically estimate the total covariance introduced from both convolution and resampling by running CWITools covariance estimation function once again. In similar fashion, we apply the scale factor to the newly created resampled variance cube.

\section{Unveiling the CGM through Deconvolution Techniques} \label{sec:algorithm}

We adopt deconvolution-based methods to separate the bright QSO core from the faint, extended \Lya\ emission in the CGM. Traditional image deconvolution techniques, such as the Richardson-Lucy (R--L) algorithm \citep{Lucy74}, provide the maximum-likelihood estimate of the true sky image given a known point-spread function (PSF). However, standard R--L suffers from noise amplification. As iterations proceed, spurious “ringing” features and pixel-level noise become increasingly enhanced. This makes it unsuitable for recovering diffuse CGM emission in the presence of a dominant point source.

To overcome these limitations, \cite{Lucy94} introduced a dual-channel deconvolution method that explicitly decomposes the image into a resolved component and an unresolved (point-source) component. This approach was originally developed for Hubble Space Telescope (HST) observations of AGN nuclei \citep{Adorf95} and later applied to IFU observations of low-redshift QSOs \citep{Fu06}. However, the substantial increase in IFU data volume and reliance on modern Python-based pipelines motivates a contemporary reimplementation of the method. Below, we first summarize the key elements of the algorithm and then present our new Python implementation, {\it CubeCarve}.

\subsection{Dual-Channel Deconvolution Framework}

The primary purpose of this algorithm is to predict the flux models of: (1) the unresolved channel representing point sources, and (2) a resolved channel describing extended structure. Prior to discussing any necessary equations, we first define some key variables that will appear throughout. All items listed below are two-dimensional arrays on an $M \times N$ pixel grid indexed by $(i,j)$. 

\begin{enumerate}

\item $\tilde{\Phi}_{i,j}$ -- The observed image

\item $\Phi_{i,j}$ -- The model image

\item $\psi_{i,j}$ -- The resolved flux channel

\item $\psi^{*}_{i,j}$ -- The unresolved flux channel

\item $\Psi_{i,j}$ -- The combined flux model $\left(\psi_{i,j} + \psi^{*}_{i,j}\right)$ 

\end{enumerate}

\noindent We adopt the the notation $\sum_{i,j}$ to denote a discrete sum over all pixel indices of an $M \times N$ array, rather than writing nested summations explicitly. 

To estimate the resolved and unresolved components, \cite{Lucy94} introduces the objective function $Q$:

\begin{equation}
Q = \mathcal{L} + \alpha S
\end{equation}

\noindent which will be maximized during image reconstruction. The quantity $\mathcal{L}$ denotes the log-likelihood, and $S$ is an entropy based regularization term where $\alpha$ controls its strength. Although \cite{Lucy94} works with the reduced log-likelihood, we adopt the full expression here:

\begin{equation}
\mathcal{L} = \sum_{i,j} \tilde{\Phi}_{i,j} \ln \Phi_{i,j} - \Phi_{i,j}
\end{equation}

\noindent The entropy term is defined in the following way:

\begin{equation}
S = - \sum_{i,j} \frac{\psi_{i,j} }{\sigma} \ln \frac{\psi_{i,j}}{\chi_{i,j}}
\end{equation}

\noindent here $\sigma$ is the total flux of the resolved component:

\begin{equation}
\sigma = \sum_{i,j} \psi_{i,j}
\end{equation}

\noindent and $\chi_{i,j}$ is the floating default. $\chi_{i,j}$ penalizes small-scale structure within the resolved channel by convolving itself with a resolution kernel $R$:

\begin{equation}
\chi_{i,j} = \sum_{m,n} \psi_{m,n}\ R_{i-m,j-n}
\end{equation}

\noindent Typically, $R$ will have a larger full width at half maximum (FWHM) than the input PSF. A broader FWHM applies a harsher penalty to unresolved structure.

Given these definitions, our goal is to determine the components $\psi_{i,j}$ and $\psi^{*}_{i,j}$ that maximizes the objective function $Q$. We perform this maximization using gradient ascent, where the update to the total flux model $\Psi$ at pixel $\left(i,j\right)$ is: 

\begin{equation}
\Delta \Psi_{i,j} = \Psi_{i,j} \frac{\partial Q}{\partial \Psi_{i,j}}
\end{equation}

\noindent The dual channel formalism then naturally separates this update into contributions from the resolved and unresolved components, allowing both channels to be updated simultaneously:

\begin{equation}
\Delta \psi_{i,j} = \psi_{i,j} \frac{\partial Q}{\partial \psi_{i,j}}\, \end{equation}

\noindent An identical expression holds for the unresolved channel by replacing $\psi_{i,j}$ with $\psi^{*}_{i,j}$. This equation shows that each pixel $(i,j)$ is updated by multiplying its current value by the local gradient of the objective function. In other words, the gradient determines the direction of change, while the current pixel value determines the scale of that change. In iterative form, the update can be written more compactly as:

\begin{equation}
\psi_{n+1} = \psi_n \left( \frac{\partial Q}{\partial \psi} \right)
\end{equation}

\noindent where $n$ and $n+1$ correspond to the current and future estimate of the chosen flux model respectively. To do this, we must derive the partial derivative term for $Q$ with respect to either channel. 

We begin with $\psi_{i,j}$. Since $Q$ is a function of $\mathcal{L}$ and $S$, we calculate the derivatives for each one independently. Referring back to equation\,2, $\mathcal{L}$ is a function of $\tilde{\Phi}_{i,j}$ and $\Phi_{i,j}$. The latter is defined as the convolution between the combined flux model and the input PSF:

\begin{equation}
\Phi_{i,j} = \sum_{m,n} P_{i-m, j-n} \left( \psi_{m,n} + \psi^{*}_{m,n} \right) 
\end{equation}

\noindent where $P_{i-m, j-n}$ is the input PSF, which we assume is normalized and invariant across the image. When we take the derivative of $\mathcal{L}$, we take it with respect to $\psi_{k,l}$. Here, the pixel $(k,l)$ represents some specific pixel location within the image. This derivative can be written as

\begin{equation}
\frac{\partial \mathcal{L}}{\partial \psi_{k,l} } = \sum_{i,j} \frac{\partial \Phi_{i,j}}{\partial \psi_{k,l}} \left( \frac{\tilde{\Phi}_{i,j}}{\Phi_{i,j}} -1 \right)
\end{equation}

\noindent The term $\frac{\partial \Phi_{i,j}}{\partial \psi_{k,l}}$ will reduce to the adjoint PSF. To show this explicitly, we start from the definition of $\Phi_{i,j}$.

\begin{equation}
\frac{\partial \Phi_{i,j}}{\partial \psi_{k,l} } = \sum_{m,n} P_{i-m,j-n} \frac{\partial }{\partial \psi_{k,l}} \left( \psi_{m,n} + \psi^{*}_{m,n}  \right)
\end{equation}

\noindent As we apply the partial derivative to $\psi^{*}_{m,n}$, we find that it will always equal 0 for any pixel location since it is independent of $\psi_{k,l}$.

\begin{equation}
\frac{\partial \psi^{*}_{m,n} }{\partial \psi_{k,l}} = 0
\end{equation}

\noindent and if we apply the same logic to $\psi_{m,n}$, the derivative will equal 0 at all pixel locations except at $(k,l)$. This gives us:

\begin{equation}
\frac{\partial \psi_{m,n} }{\partial \psi_{k,l} } = \delta_{mk} \delta_{nl}
\end{equation}

\noindent Now, we can rewrite Equation\,11 as:

\begin{equation}
\frac{\partial \Phi_{i,j} }{\partial \psi_{k,l} } = \sum_{m,n} P_{i-m,j-n} \delta_{mk} \delta_{nl}
\end{equation}

\noindent and we remove the $\delta$ terms by plugging in the only indices that prevents the term from equaling zero:

\begin{equation}
\frac{\partial \Phi_{i,j} }{\partial \psi_{k,l} } = P_{i-k, j-l}
\end{equation}

\noindent We now plug in Equation\,15 into Equation\,10:

\begin{equation}
\frac{\partial \mathcal{L}}{\partial \psi_{k,l} } = \sum_{i,j} P_{i-k, j-l} \left( \frac{\tilde{\Phi}_{i,j}}{\Phi_{i,j}} - 1 \right) 
\end{equation}

\noindent Note that $P_{i-k, j-l}$ appears with reversed indices inside the summation. This is simply the adjoint (i.e., flipped) PSF, which we denote by $P^{\dagger}$. Using this notation, we can rewrite Equation\,16 in a compact convolution form:

\begin{equation}
\frac{\partial \mathcal{L}}{\partial \psi_{k,l}} = \left[ P^{\dagger} \ast  \left( \frac{\tilde{\Phi}}{\Phi } \right) - P^{\dagger} \ast 1 \right]_{k,l}
\end{equation}

\noindent where $\ast$ denotes a convolution. Because we adopt a PSF normalized such that $\sum_{i,j} P_{i,j} = 1$, the convolution $P^{\dagger} \ast 1$ returns a constant array equal to 1 everywhere. This constant can be absorbed into the overall normalization of the gradient update and therefore does not affect the multiplicative correction factor. For this reason, the $P^{\dagger} \ast 1 $ term can be dropped. For readers familiar with standard R--L deconvolution, Equation\,16 becomes the R--L Correction factor ($C_{\rm RL}$):

\begin{equation}
\left[ P^{\dagger} \ast  \left( \frac{\tilde{\Phi}}{\Phi } \right) \right]_{k,l} \equiv C_{\rm RL}
\end{equation}

We now continue on to the partial derivative of the entropy term using Equation\,3. As a reminder, the $\sigma$ and $\chi_{i,j}$ terms act as constants, so they do not have any impact on the derivative term. We take the partial derivative with respect to $\psi_{k,l}$ again.

\begin{equation}
\frac{\partial S}{\partial \psi_{k,l}} = \sum_{i,j} \frac{-1}{\sigma} \left( \ln \left[ \frac{\psi_{i,j}}{\chi_{i,j}} \right] + 1 \right) \delta_{ik} \delta_{jl}
\end{equation}

\noindent As before, the $\delta_{ik} \delta_{jm}$ terms indicate that the expression will be zero unless the pixel locations equal $(k,l)$. 

\begin{equation}
\frac{\partial S}{\partial \psi_{k,l}} = \frac{-1}{\sigma} \left( \ln \left[ \frac{\psi_{k,l}}{\chi_{k,l}} \right] + 1 \right)
\end{equation}

With both partial derivative terms derived, we can combine them together to form our multiplicative update as written in Equation\,8.

\begin{equation}
\psi_{n+1} = \psi_{n} \left( \frac{\partial \mathcal{L}}{\partial \psi} + \alpha \frac{\partial S}{\partial \psi} \right)
\end{equation} 

\noindent We drop the pixel indices as the operation acts on the full 2D array. This can be rewritten into a clean form by plugging in each partial derivative term:

\begin{equation}
\psi_{n+1} = \psi_n \left( C_{\rm RL} + \frac{-\alpha}{\sigma} \left( \ln \left[ \frac{\psi}{\chi} \right] + 1 \right) \right)
\end{equation}

With the resolved component complete, we finish with the unresolved component. Since it follows the same procedure as before, we reuse Equation\,21. However, $S$ has no dependence on $\psi^{*}$, so $\frac{\partial S}{\partial \psi^{*}} = 0$. As a result, we are left with the simple multiplicative update for the unresolved channel:

\begin{equation}
\psi^{*}_{n+1} = \psi^{*}_{n} \, C_{\rm RL}
\end{equation}

\noindent which is the standard R--L method from \cite{Lucy74}. 

\subsection{Implementing Dual-Channel with {\it CubeCarve}}
We are now ready to continue on to the modern implementation of this algorithm with {\it CubeCarve}, which is publicly available on GitHub\footnote{\url{https://github.com/kevhall23/CubeCarve}}. The core algorithm follows the methodology discussed in the previous section but within a Python framework. {\it CubeCarve} is intended to operate on IFU data, but we will keep the focus of this discussion on 2D images and extend to 3D in the next section.

Given some input image that contains a combination of unresolved point sources (e.g., stars or QSOs) and extended gaseous emission (e.g., CGM emission), the user must first provide initial guesses on the locations for all point sources. To do this, a 2D NumPy array, matching the shape of the input image, can be made where all pixels are either one or zero. The pixels representing point sources are assigned the value of one. 

Since the point sources are likely to fall at sub-pixel locations, {\it CubeCarve} supersamples the input image by a scale factor $\kappa$. This is achieved by using Scipy's bicubic interpolator, \texttt{RectBivariateSpline}. To preserve flux, we scale the upsampled image by $1/\kappa^2$. With our supersampled image and initial point source array at hand, we perform 2D Gaussian fits on the supersampled image to update the initial guesses. For our initial centroid position, we simply scale the non-zero pixel location in the input array by $\kappa$. For example, if $\kappa=10$ and an initial point source location is $(2,2)$, the updated location on the supersampled image is $(20,20)$. {\it CubeCarve} will automatically perform these fits for each declared point source and save their best-fit positions within a new supersampled 2D array matching the shape of the supersized image. 

This array of best-fit point source positions on the upsampled grid corresponds to the initial unresolved flux model $\psi^{*}$. In the original \cite{Lucy94} implementation, the $\psi^*$ array maintains this format, where the point sources are fixed delta functions through each multiplicative update. In {\it CubeCarve}, rather than single stationary pixel location for each point source, we place a 2D Gaussian at each location. The 2D Gaussian has a FWHM equal to the input PSF, which is empirically estimated. The details on how the empirical PSF model is built will be discussed in the next section. For now, we assume that the PSF is known. This modification allows the unresolved flux to dynamically update its position and brightness within neighboring pixels through each iteration. 

Once the unresolved channel is built, {\it CubeCarve} constructs the initial resolved channel $\psi$. We simply create an array matching the shape of the supersampled image where each pixel has a value of 1. With both the unresolved and resolved channels built, we specify that both have equal flux prior to running the algorithm. Specifically, we require:

\begin{equation}
\sum_{i,j} \psi_{i,j} + \psi^{*}_{i,j} = 1
\end{equation} 

\noindent This requirement avoids any channel imbalance and ensures fair distribution of flux between them. Next, we estimate the noise level of the image. For this, {\it CubeCarve} allows the user to supply a region within the image absent of any obvious emission. Rather than calculating the root-mean-square (RMS), we just calculate the mean pixel value within this region. We use this value as an additional weight term in the update term for the resolved channel. Specifically, we modify Equation\,22 in the following way:

\begin{equation}
\psi_{n+1} = \psi_n \left( C_{\rm RL} + W^2 \frac{-\alpha}{\sigma} \left( \ln \left[ \frac{\psi}{\chi} \right] + 1 \right) \right)
\end{equation}

\noindent where $W$ is the additional weight term derived from the background level. We square this value to ensure it has a larger impact on the iterative update and to avoid negative values.

From this point, {\it CubeCarve} follows the procedure discussed in the previous section. At the beginning of each iteration, the model image $\Phi_{i,j}$ is built and scaled to the flux of the input image $\tilde{\Phi}_{i,j}$. Next, the update procedures are performed for each flux channel respectively. After each update, we apply a flux correction $\lambda$ on to each channel, which we define as:

\begin{equation}
\lambda = \sum_{i,j} \frac{\tilde{\Phi}_{i,j}}{\Psi_{i,j} + \epsilon}
\end{equation}

\noindent with $\epsilon \ll 1$ to avoid any divisions of zero in the ratio. This factor prevents either channel from diverging radically from a realistic solution. Once $\lambda$ is multiplied to each channel, the next iteration begins.

{\it CubeCarve} requires a maximum number of iterations $(N)$ to run by the user. However, the algorithm terminates once the model finds a stable solution. Convergence is assessed by computing the fractional change in total flux for the resolved array at the end of each iteration. When $\Delta\psi < 10^{-3}$, iterations stop. 

\DC\ outputs two products: (1) The resolved model, which represents the final extended emission map, downsampled to the original spatial resolution. (2) The unresolved model, which has been convolved with the empirically estimated PSF and downsampled to the original spatial resolution. Because $\psi$ directly represents the extended emission with the QSO entirely excluded, no explicit PSF subtraction is required. \DC\ effectively “carves out” the resolved emission, eliminating any possible subtraction artifacts. Meanwhile, $\psi^{*}$ provides stable, wavelength-dependent flux estimates for the point sources, enabling highly accurate PSF scaling if desired.

\subsection{Deployment of {\it CubeCarve}}

Once our KCWI and resampled MUSE datacubes are assembled, we are ready to extract the extended \Lya\ emission from each SMG-QSO target with {\it CubeCarve}. To process a full datacube, we run {\it CubeCarve} iteratively on every wavelength slice. Each 2D slice requires a PSF model for the dual-channel deconvolution. We construct empirical PSF models following methods used in previous QSO studies \citep[e.g.,][]{Borisova16,Arrigoni-Battaia19,Cai19,OSullivan20a,Gonzalez-Lobos23}, with a few notable differences summarized below.

After the user identified point sources are fit on to the supersampled grid, rectangular cutouts are placed around them. These cutouts define the spatial extent of the empirical PSF used in the deconvolution. The user can choose the PSF size; in this work we adopt a $4\farcs0 \times 4\farcs0$ region. For each wavelength slice, we construct the empirical PSF by combining 200 neighboring wavelength layers with a sigma-clipped mean, yielding a high-S/N model. Layers within a $\pm3000$\,\kms\ window around the \Lya\ line are excluded from the model to prevent contamination. If multiple point sources are present, {\it CubeCarve} averages their individual cutouts to produce a single model.

In practice, we initialize two empty 3D arrays: one for the resolved-emission model and one for the PSF-subtracted (unresolved) model. For each wavelength slice, we compute both the resolved and unresolved flux contributions and store the results in their respective arrays. Convergence is typically achieved within $N\sim 50-100$ iterations using a regularization strength of $\alpha = 1\times10^{-4}$. We also adopt a resolution kernel $R$ whose FWHM is set to five times that of the empirical PSF. Because our goal is to study the extended emission, not the deconvolved image, we reconvolve the resolved component with the empirical PSF before writing it to the cube. For the unresolved component, we scale the empirical PSF by the fitted point source flux and subtract it from the data. To minimize any subtraction artifacts, all extended \Lya\ measurements are drawn exclusively from the resolved-emission model.

Finally, to remove the remaining continuum sources, we apply a running-median continuum subtraction. A median over 300 adjacent wavelength channels provides a robust continuum estimate for each spectral slice, enabling reliable subtraction.
\subsection{Validation and Performance Tests}

Prior to deploying {\it CubeCarve} on our KCWI and resampled MUSE data, we conducted a series of experiments using controlled “pseudo” datacubes. These experiments allowed us to test the algorithm’s performance under realistic yet fully controlled conditions. To construct one of these test cubes, we extracted a spectrum from the Sloan Digital Sky Survey Data Release 16 Quasar catalog \citep[SDSS DR16Q;][]{Lyke20} to serve as the target in our simulated  observation. We selected QSO\,J0050+0051 at $z=2.219$, which was previously observed by KCWI \citep{Cai19}. The wavelength range was truncated to match the approximate spectral coverage of the KCWI blue channel. No resampling was performed to reproduce the exact KCWI wavelength sampling, as this does not affect {\it CubeCarve}’s performance.

We structured the mock datacube to resemble a single, fully reduced KCWI exposure. The spatial plate scale matches the large image slicer configuration, such that each spaxel corresponds to approximately 1\farcs35\,$\times$\,0\farcs29. The QSO is placed at the center of the field, and an extended emission region representing the \Lya\ nebula is added as a circular mask centered on the QSO with a radius of $\approx$\,4\arcsec. Due to the rectangular spaxel shape, the nebula appears slightly elongated along the $y$-direction. The \Lya\ emission line is modeled with a 1D Gaussian centered at the redshifted \Lya\ wavelength and a FWHM of $\approx$\,300\,\kms. To keep the example simple, no velocity gradients or kinematic substructure are introduced.

\begin{figure}[!htb]
\centering
\includegraphics[width=0.45\textwidth]{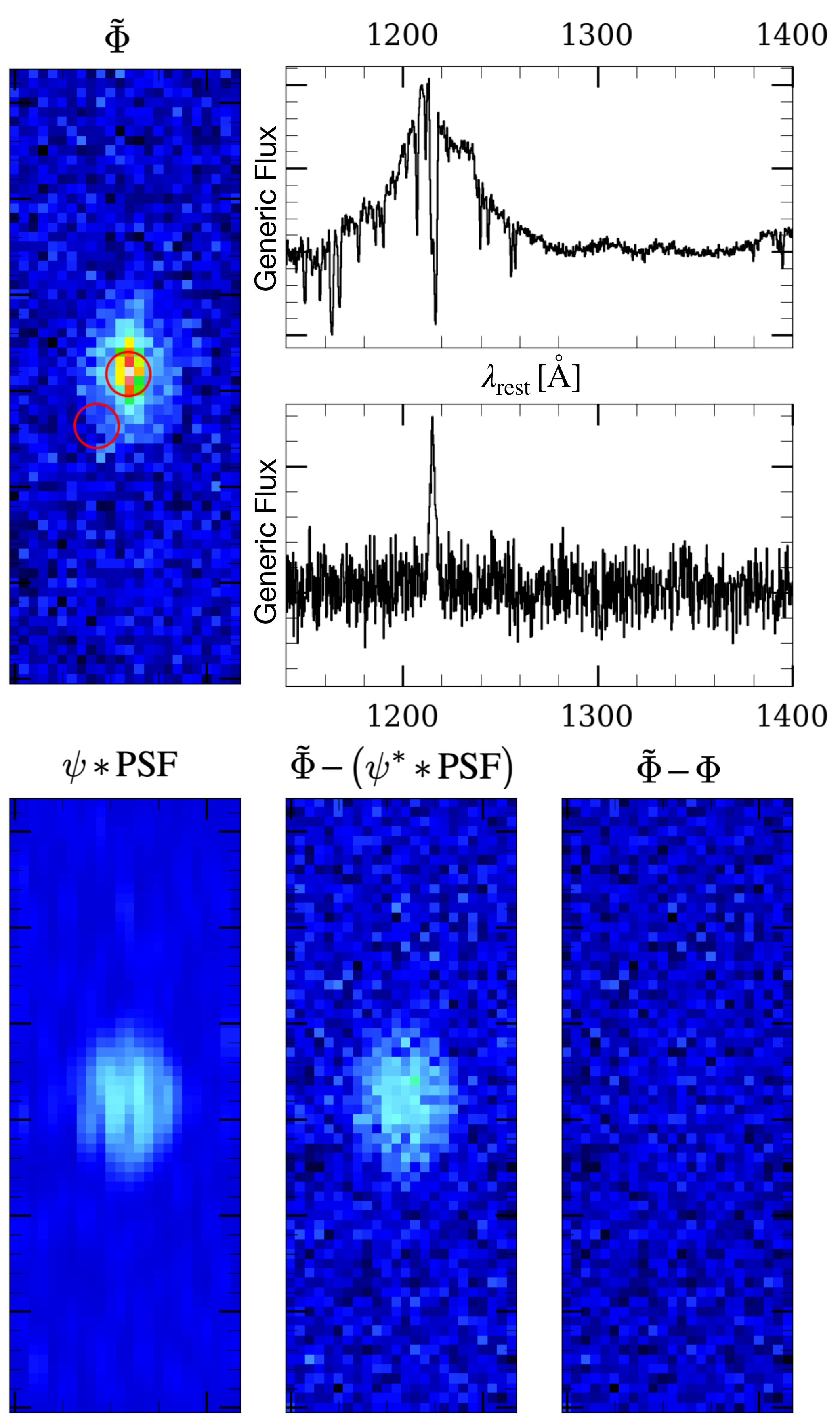}
\caption{A visual demonstration of {\it CubeCarve} deployed on a simulated 3D datacube. {\it Top-Left}: The ``observed'' image taken from a single wavelength slice. {\it Top-Right}: 1D spectra of different spatial locations within the datacube. {\it Bottom}: From left to right, the resolved emission model, PSF subtraction using the inferred flux from the unresolved emission model, and the residual between the observed image and the final model. }
\label{fig:validation-figure}
\end{figure}

Once all components are assembled, each wavelength slice is convolved with a realistic KCWI PSF. Although the PSF in real data is wavelength-dependent, here we adopt a representative PSF derived from previous KCWI exposures. Finally, a Gaussian noise model is applied to every layer of the datacube to mimic observational noise. The top-left panel of Figure\,\ref{fig:validation-figure} shows a single wavelength slice where the simulated \Lya\ nebula is visible. The top-right panel presents representative 1D spectra extracted from the regions marked in red apertures with one centered on the bright QSO and another sampling the faint, extended emission.

The bottom panels of Figure\,\ref{fig:validation-figure} present the results of applying {\it CubeCarve} to this simulated dataset. The algorithm successfully separates the unresolved and resolved emission components, yielding a clean reconstruction of the faint extended signal. The combined model, $\Phi$, produces a flat residual image with no significant structure or artifacts. These tests demonstrate that {\it CubeCarve} accurately recovers both compact and diffuse emission, validating its reliability for application to real IFU observations. 

\begin{figure*}[!htb]
\includegraphics[width=\textwidth]{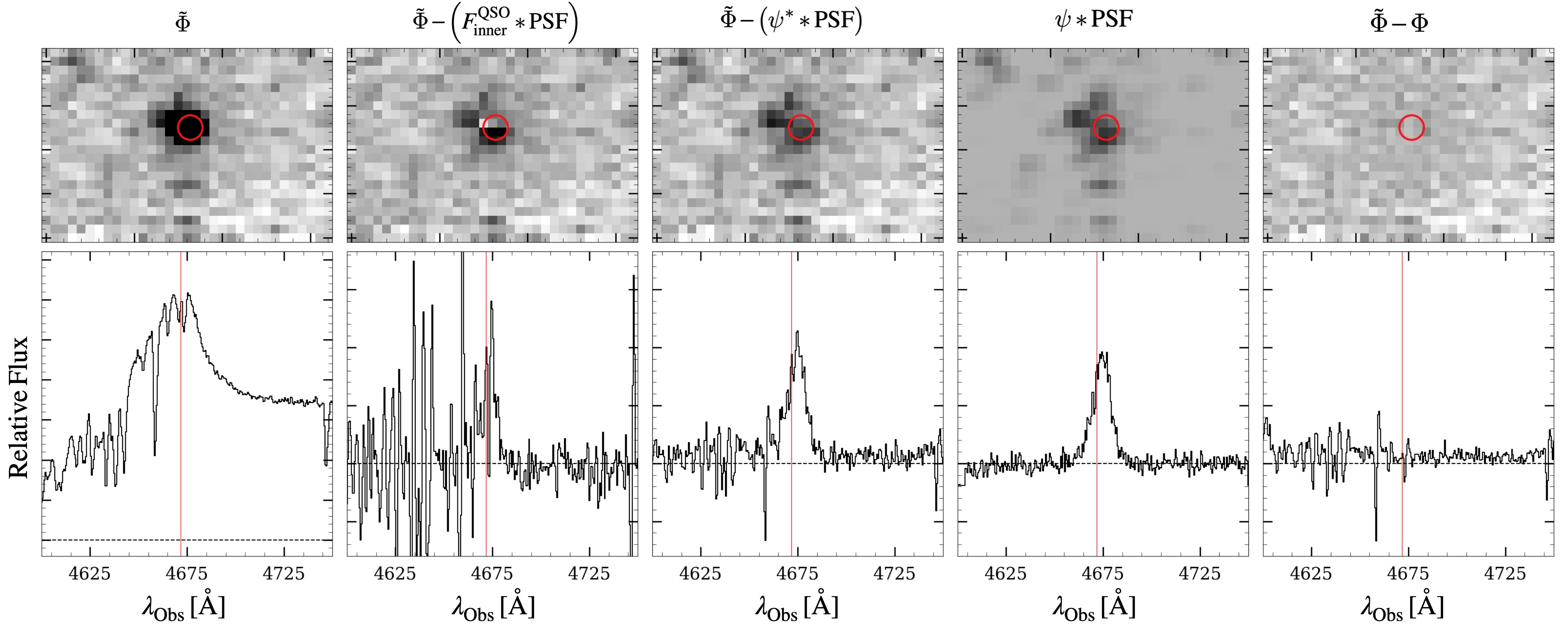}
\caption{ {\it Top Row}: \Lya\ SB map for a single wavelength layer containing the nebula. {\it Bottom Row}: Integrated spectrum extracted from the red aperture shown in the top panels. The vertical red line marks the chosen wavelength layer used in the top row. }
\label{fig:DC-compare}
\end{figure*}

To better understand the performance of {\it CubeCarve} on real IFU data, we perform a comparison against the traditional PSF subtraction method commonly used in the literature \citep[e.g.,][]{Borisova16,Arrigoni-Battaia19}. Specifically, we deploy the analysis tool developed and used in \citetalias{Hall24} for the GAMA\,J09013$-$0107 system. We previously used the "direct approach" method, which scales the empirical PSF by the "inner" (e.g., $\sim 1-2"$) flux of the QSO at a given wavelength slice. This can be mathematically understood as $\tilde{\Phi} - \left( F_{\rm inner}^{\rm QSO} \ast \rm PSF \right)$.  For this comparison, we selected \idTwo\ and present the results of this comparison in Figure\,\ref{fig:DC-compare}. By quick inspection, it is easy to spot the over subtraction residual at the location of the QSO using the direct approach method (second column in Figure\,\ref{fig:DC-compare}). Contrast that with the output from {\it CubeCarve}, there are no residuals remaining. 

We continue this comparison by extracting 1D spectra by integrating within the red apertures shown in each panel in the top row. The spectrum from the direct approach method really emphasizes why the inner regions of the QSO must be masked prior to emission extraction. The strong residuals make it incredibly difficult to isolate emission from noise. Furthermore, we lose the true signature of the emission line itself. However, we observe a clean emission line in both data products from {\it CubeCarve}. The subtraction method using the unresolved flux model does have a few artifacts, but they are much less severe. 

The primary takeaway from this exercise is the impact {\it CubeCarve} can have on CGM studies. It delivers a clean view of the extended emission surrounding the QSO. Most importantly, we can now study the inner regions of the nebulae for QSO systems. This can have significant impacts on what we can extract from the \Lya\ nebula, which we will explore in the following section. As shown here, {\it CubeCarve} can work for both KCWI and MUSE, and it has the capability to operate on even more instruments in the near future. 

\section{Results}\label{sec:results}

\begin{figure}[!htb]
    \centering
    \hspace*{-0.4cm}
    \includegraphics[width=0.5\textwidth]{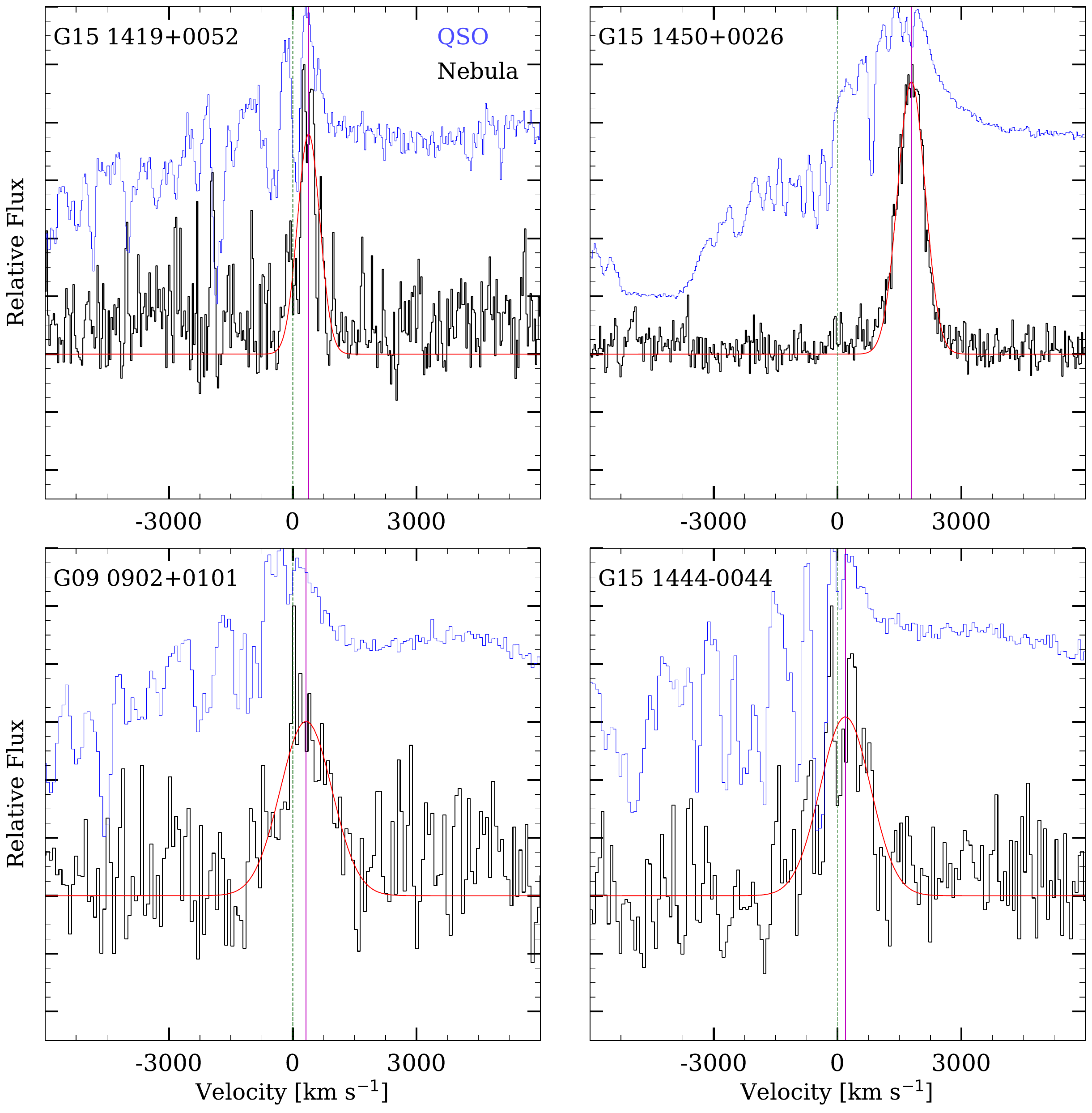}
    \caption{The normalized flux density for both the QSO and the \Lya\ nebula as a function of velocity. For all panels, we plot the spectrum of the QSO and \Lya\ nebula for each \SQSO\ composite system as blue and black lines respectively and overlay the best-fit 1D Gaussian in red with its mean value shown as a vertical magenta line. The green dashed-line corresponds to the \Lya\ line at the redshift of each galaxy. }
    \label{fig:neb-vs-qso-spectra}
\end{figure}

\begin{figure*}[!htb]
    \centering
    \includegraphics[width=\textwidth]{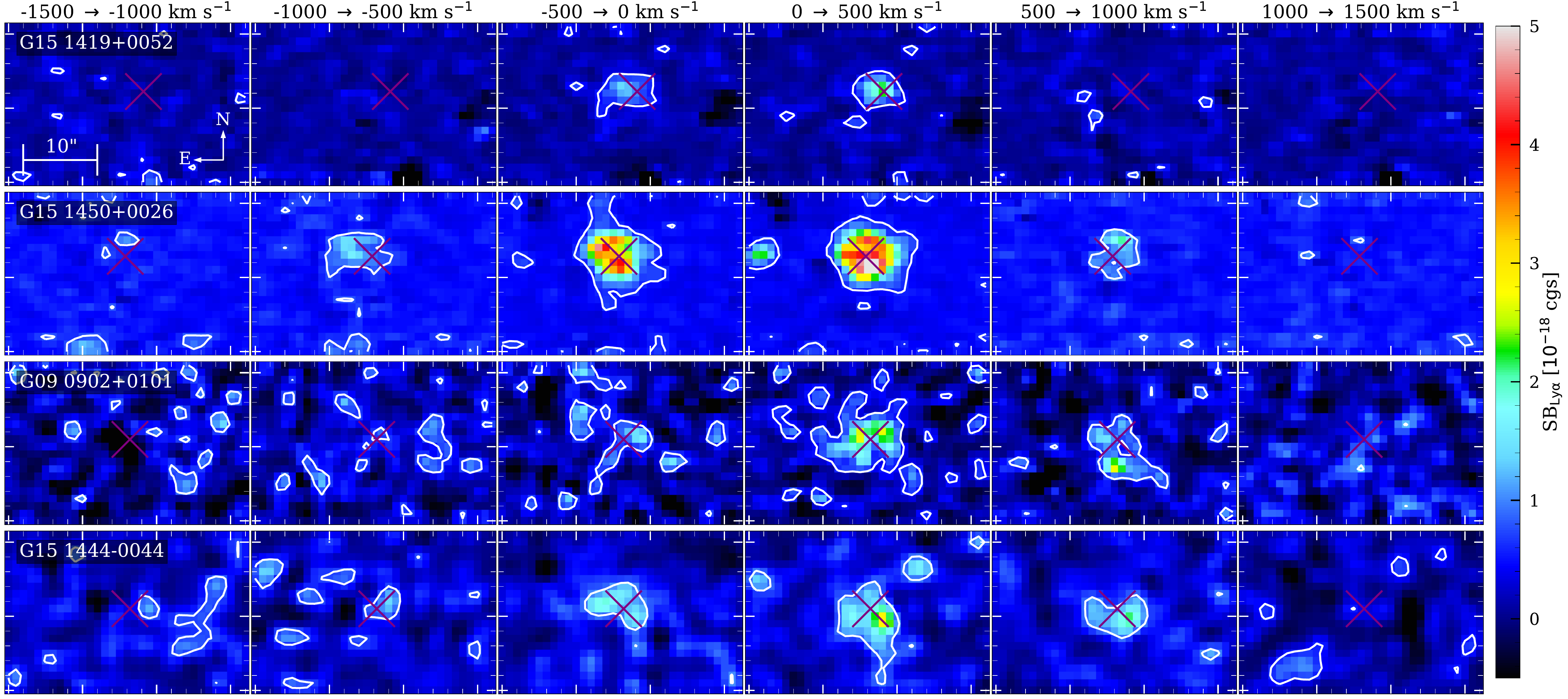}
    \caption{\Lya\ channel maps. We select a total velocity range of $\pm 1500$\,\kms. Each column corresponds to a narrow $\approx 500$\,\kms\ channel for each SMG-QSO, as denoted at the top of each column. The top two rows correspond to the KCWI targets, and the bottom two rows correspond to the MUSE targets. The white contour corresponds to S/N = 2. For presentation purposes, we convolve each channel map with a $2\times2$ Boxcar kernel. The purple X corresponds to the position of the QSO. }
    \label{fig:channel-maps}
\end{figure*}

We are now ready to study the extended \Lya\ line emission for all four \SQSO\ composite systems. In this section, we focus on the measured properties of these nebulae. Here, we use the resolved emission output ($\psi$) from {\it CubeCarve} to measure the \Lya\ emission properties. We begin our characterization of the extended \Lya\ emission by extracting 1D spectra. In doing so, we can ensure no remaining QSO emission is present within the datacube. We extract the flux within a $4\farcs5$ aperture centered on the optical position of each QSO from both the original and resolved emission only datacubes. Figure \ref{fig:neb-vs-qso-spectra} presents the spectra, as well as the best-fit 1D Gaussian. Overall, there is no presence of the QSO remaining in the datacube. The dotted green line corresponds to $z_{\rm QSO}$ and the dotted magenta line corresponds to the redshift of the \Lya\ nebula ($z_{\rm Ly\alpha}$). We report $z_{\rm Ly\alpha}$ for each target in Table\,\ref{tab:lya_results}, and it will be the value used in reference to future velocity measurements. We find that $z_{\rm Ly\alpha}$ is near $z_{\rm QSO}$ for all targets except for \idTwo. There is a $\approx 1800$\,\kms\ velocity offset between the nebula redshift and $z_{\rm QSO}$. This is an abnormally large velocity offset, so it may be the case of an incorrect redshift identification for \idTwo.

We continue our characterization of the emission for each \SQSO\ target by producing \Lya\ channel maps. For each target, we select a $\pm 1500$\,\kms\ window with respect to $z_{\rm Ly\alpha}$ and produce a series of narrow-band images each with a width of $\approx 500$\,\kms\ (Figure\,\ref{fig:channel-maps}). We generate S/N $> 2$ contours by estimating the $1\sigma$ noise level by taking the RMS from regions absent of obvious \Lya\ emission. Each \SQSO\ composite system hosts extended \Lya\ emission spanning a wide velocity range. We can now begin to identify differences in \Lya\ nebula characteristics between our targets.

Starting at the lowest redshift within our sample, \idOne\ does not exhibit broad \Lya\ emission. We only detect significant emission within the $\delta v \approx 0 \rightarrow 500$\,\kms\ channel. Furthermore, we do not detect any \Lya\ emission further than $\sim 5\farcs0$ from the position of the galaxy. This is in stark contrast to the \idTwo, which possesses noticeably broader \Lya\ emission. There is clear emission spanning $\delta v \approx -500 \rightarrow 1000$\,\kms. Additionally, the \Lya\ SB is significantly brighter than what we observe in \idOne. Interestingly, we detect a small region roughly $10\farcs0$ to the east of the QSO position. This source is interestingly not seen in the deep KiDs image. 

Moving on to the \SQSO\ targets at $z > 3$, our \Lya\ channel maps will suffer from reduced S/N due to our MUSE resampling procedure which added additional noise. Nonetheless, we still detect clear \Lya\ emission within both targets, which matches our expectations following the results from \cite{Gonzalez-Lobos23}. For \idThree, the \Lya\ emission is relatively narrow, as we only find significant detections within the $\delta v \approx 0 \rightarrow 500$\,\kms\ channel. However, \idFour\ exhibits both brighter and broader ($\delta v \approx -500 \rightarrow 1000$\,\kms) \Lya\ emission. There are a few isolated pockets of emission outside this window, but they are likely spurious detections. 

From the \Lya\ channel maps, we establish that all four galaxies possess extended \Lya\ nebula. Unsurprisingly, each nebula has unique characteristics. However, this quick examination has revealed a possible pattern. Both \idTwo\ and \idFour\ share the broadest and brightest \Lya\ emission, while \idOne\ and \idTwo\ exhibit narrower and fainter emission. Is this a real pattern that signals a difference in their environment? To answer this question, we seek a closer inspection into each galaxy by producing high S/N moment maps and measuring their \Lya\ nebula properties.

\subsection{\Lya\ Moment Maps} \label{subs:lya-moment-maps}

\begin{figure*}[!htb]
    \centering
    \includegraphics[width=\textwidth]{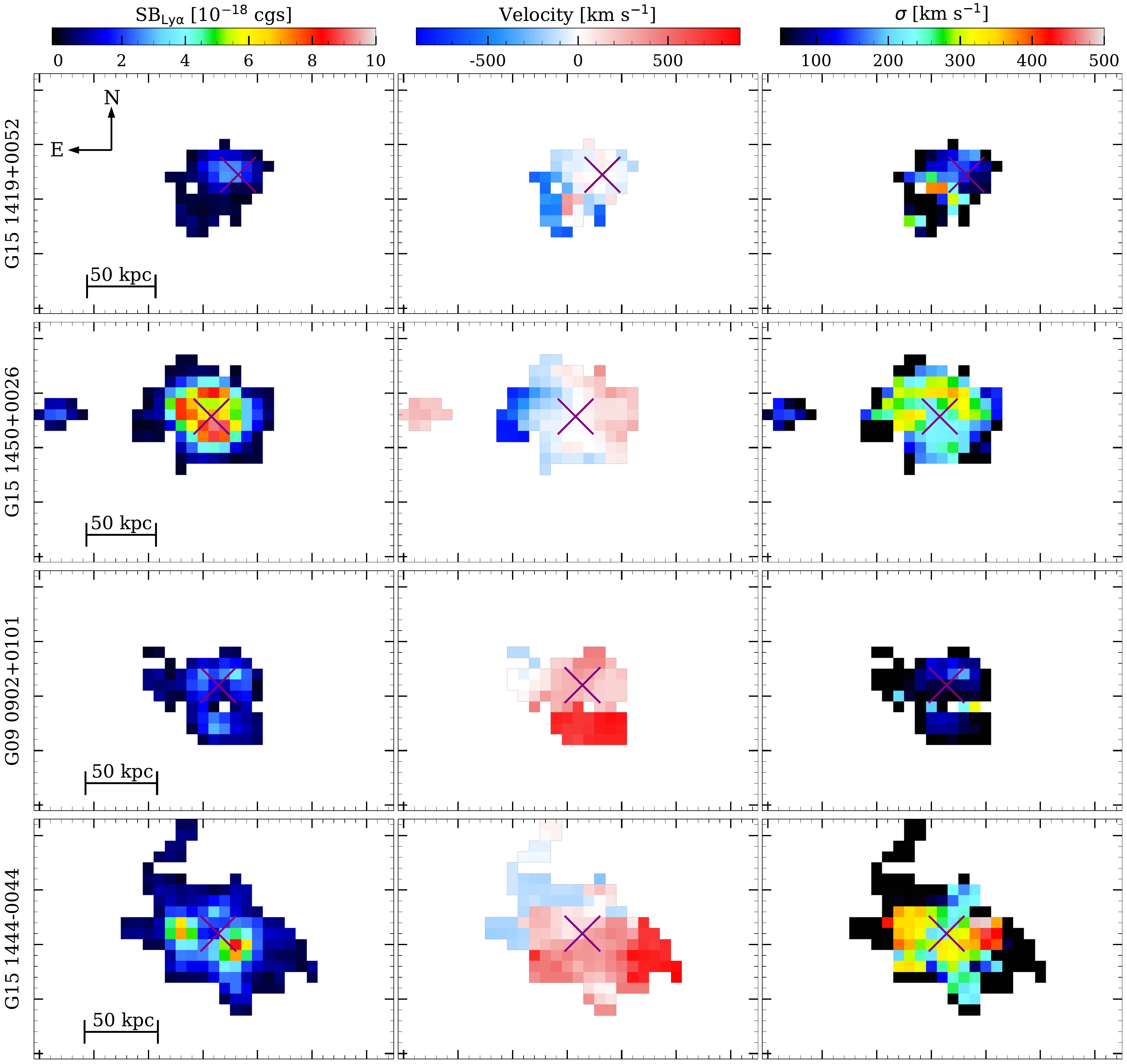}
    \caption{Moment maps of the extended \Lya\ emission surrounding the \SQSO\ composite systems. \idOne, \idTwo, \idThree, \idFour\ from top to bottom, and the \Lya\ surface brightness, velocity, and dispersion maps from left to right. All targets share an equal colorbar for their respective moment maps, which can be found at the top of each column. We provide a scale bar set to 50 kpc at $z_{\rm QSO}$ (see Table\,\ref{tab:lya_results}) for each target, in the lower left corner of each moment 0 map.  }
    \label{fig:moment-maps}
\end{figure*}

We construct \Lya\ moment maps to obtain a comprehensive view of each nebula’s morphology and kinematics. To define the spatial extent over which moments are computed, we first collapse each datacube over the spectral range containing the \Lya\ emission line using our 1D Gaussian fits seen in Figure\,\ref{fig:neb-vs-qso-spectra} and generate a S/N $> 2$ contour; this contour establishes the nebula boundary used for the moment maps. Figure\,\ref{fig:moment-maps} presents the resulting zeroth, first, and second moments, corresponding to the \Lya\ surface brightness, flux-weighted velocity, and velocity dispersion, respectively. From these maps, we measure the total \Lya\ luminosity ($L_{\rm Ly\alpha}$), projected nebula area, and peak surface brightness (SB$_{\rm peak}$), as summarized in Table\,\ref{tab:lya_results}. In the following, we describe each system in order of increasing redshift.

The \Lya\ emission from \idOne\ forms a compact halo centered on the QSO. Consistent with the channel maps in Figure \ref{fig:channel-maps}, we detect no significant \Lya\ flux beyond $\sim$5\arcsec\ ($\approx40$\,kpc at $z=2.67$). The nebula peaks at SB$_{\rm peak}=2.8\times10^{-18}$\,\surf\ and has an average brightness of SB$_{\rm avg}=2.2\times10^{-18}$\,\surf. Integrating over the full 3D aperture yields $L_{\rm Ly\alpha}=0.6\pm0.15\times10^{43}$\,erg\,s$^{-1}$. The velocity field suggests a modest gradient from southeast (blueshifted) to northwest (redshifted), though the evidence remains tentative. The velocity dispersion is relatively uniform, with a mean value of $\sigma_v \sim$200\,\kms\ and no evidence for strong turbulence within the halo.

\begin{table*}[!htb]
        \caption{Properties of Extended \Lya\ Nebulae}
    \label{tab:lya_results}
   \hspace*{-0.3cm}
    \begin{tabular}{lccccrl}
        \hline
        \hline
        Parameter & \idOne & \idTwo & \idThree & \idFour & Unit &  \\
        \hline
        \multicolumn{7}{l}{ {\bf QSO Properties} } \\
        $z_{\rm QSO}$ & 2.6711 & 2.8220 & 3.1204 & 3.3750 & --- & (1) \\
        $L_{\rm bol}$ & $46.2\pm0.2$ & $46.7\pm0.2$ & $46.6\pm0.2$ & $46.6\pm0.2$ & $\log (\rm erg\,s^{-1})$ & (2) \\
        $M_{\rm BH}$ & $9.0\pm 0.4$ & $9.1\pm0.4$ & $9.1\pm0.4$ & $9.3\pm0.4$ & $\log (M_{\odot})$ & (3) \\
        $\dot{M}_{\rm BH}$ & 2.3 & 7.1 & 5.5 & 6.2 & \msunyr & (4) \\
        $\lambda_{\rm Edd}$ & 0.22 & 0.12 & 0.18 & 0.30 & --- & (5) \\
        \hline 
        \multicolumn{7}{l}{ {\bf Host Galaxy Properties}} \\
        
        $S_{870}$ & $6.4\pm0.6$ & $3.0\pm0.8$ & $14.4\pm0.8$ & $6.0\pm0.6$ & mJy & (6) \\
        $L_{\rm IR}^{\rm SFR}$ & $13.0\pm0.2$ & $12.9\pm0.2$ & $13.2\pm0.2$ & $13.3\pm0.2$ &  $\log (L_{\odot})$ & (7) \\
        $M_{\rm gas}$ & $11.58\pm0.12$ & $11.04\pm0.16$ & $11.95\pm0.12$ & $11.56\pm0.12$ & $\log (M_{\odot})$ & (8) \\
        SFR & 1000 & 800 & 1700 & 1900 & \msunyr & (9) \\
        \hline 
         \multicolumn{7}{l}{{\bf CGM Properties}} \\
        SB Limit & 1.7 & 1.6 & 2.9 & 2.9 & $\times 10^{-18}$\,\surf & (10) \\
        $z_{\rm Ly\alpha}$ & $2.6758\pm0.0002 $ & $2.8449\pm0.0001 $ & $3.1248\pm 0.0019$  & $3.3778\pm0.0022 $ & --- & (11) \\
        Luminosity & $0.60 \pm 0.15$ & $ 2.5 \pm 0.13$ & $1.4 \pm 0.10$ & $2.1 \pm 0.15$ & $10^{43}$\,\ergps & (12) \\
        Area       & 30 (1894) & 92 (5625) & 69 (3995) & 140 (7697)  & arcsec$^2$ (kpc$^2$) & (13) \\
        SB$_{\rm peak}$ & 2.8 & 9.2 & 3.6 & 8.4 & $\times 10^{-18}$ \surf & (14)  \\
        \hline
        \hline
        & & & & & & \\
        
    \end{tabular}
    \tablecomments{  Rows: (1) Redshift of QSO.  (2) QSO bolometric luminosity derived from rest-frame 1350\,\AA\ luminosity \citep{Hopkins07}. (3) Virial Black Hole mass estimated by C\,{\sc{iv}}. (4) Accretion rate using $L_{\rm bol}^{\rm QSO}$ and a radiative efficiency of $\epsilon = 0.1$. (5) Eddington ratio of the QSOs. (6) ALMA 870\,\um\ flux density. (7) Rest-frame 8-1000\,\um\ luminosity powered by star formation.  (8) Gas Mass from ALMA 870\,\um\ photometry. (9) SFR derived from $L^{\rm SFR}_{\rm IR}$. (10) The $2\sigma$ surface brightness limit in a $4\farcs0$ aperture within $\pm 1500$\,\kms\ centered at the \Lya\ line of each target. (11) Redshift of \Lya\ nebula taken from spectra presented in Figure \ref{fig:neb-vs-qso-spectra}, (12-14) \Lya\ characteristics for each SMG-QSO target.  }
\end{table*}

In contrast, \idTwo\ exhibits markedly brighter and more extended emission. Its peak surface brightness reaches $9.2\times10^{-18}$\,\surf, roughly three times higher than \idOne. The integrated luminosity is $L_{\rm Ly\alpha}=2.5\pm0.13\times10^{43}$\,erg\,s$^{-1}$. The velocity map reveals a distinct east-west gradient, with blueshifted gas to the east and redshifted gas to the west, indicative of possible large-scale outflows. The velocity dispersion is elevated compared to \idOne, reaching up to 400\,\kms, consistent with a more dynamic and disturbed CGM environment. The compact emission knot, previously detected in the channel map, at $\delta v\approx400$ km s$^{-1}$ may trace either a dense CGM cloud or a faint satellite galaxy below the continuum detection limit of the deep KiDS imaging. It's relative velocity is consistent with material gravitationally bound to the halo hosting \idTwo. 

The morphology of \idThree\ more closely resembles that of \idOne. It is a compact nebula with no significant emission detected beyond $\sim35$\,kpc. The peak and average SBs are modest (SB$_{\rm peak}=3.6\times10^{-18}$\,\surf, and the integrated luminosity $L_{\rm Ly\alpha}=1.4\pm0.10\times10^{43}$\,erg\,s$^{-1}$ is roughly half the value reported by \cite{Gonzalez-Lobos23}. The velocity field shows no evidence of coherent gradients, and the velocity dispersion remains moderate ($\sigma_v \approx 300$\,\kms). The lower dispersions relative to previous results may reflect the impact of our resampling procedure, which can systematically reduce the apparent line width.

Our highest-redshift source, \idFour, displays a larger spatial extent and enhanced \Lya\ brightness near the QSO position. The nebula peaks at SB$_{\rm peak}=8.4\times10^{-18}$\,\surf\ and has a total luminosity of $L_{\rm Ly\alpha}=2.1\pm0.15\times10^{43}$\,erg\,s$^{-1}$. As with \idThree, this value is lower than previously reported by \cite{Gonzalez-Lobos23}. The velocity map, however, reveals clear kinematic structure. Redshifted gas ($\sim 300$\,\kms) dominates the southern portion of the nebula, transitioning to blueshifted velocities toward the north. The dispersion map shows the highest velocity widths among our four \SQSO\ systems, reaching $\sim 450$\,\kms, suggesting an especially energetic or turbulent CGM.

\section{Discussion}\label{sec:comparisons}

\subsection{Azimuthally Averaged Radial Profiles} \label{subs:lya-radial-profs}

\begin{figure*}[!htb]
    \centering
    \includegraphics[width=\textwidth]{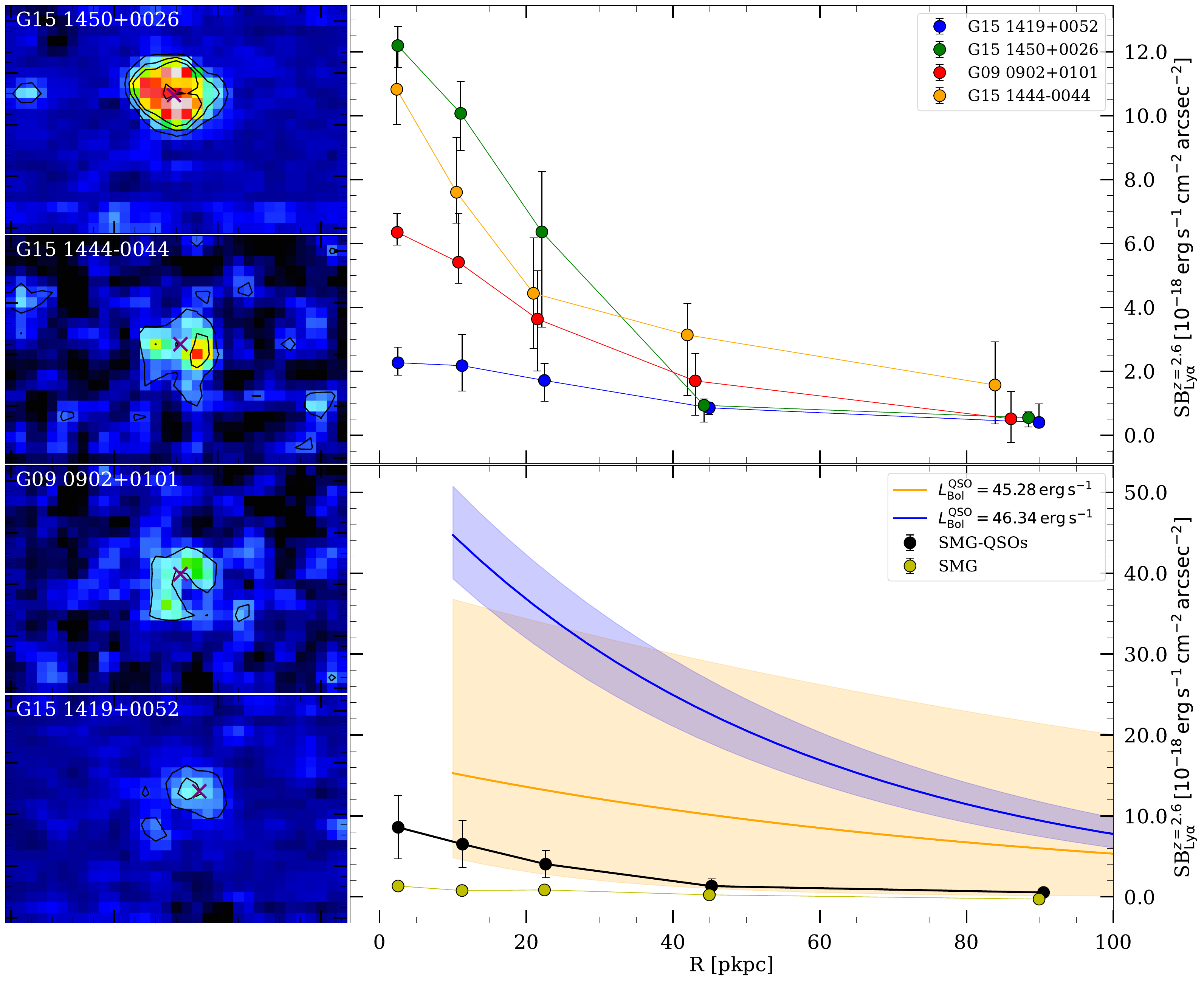}
    \caption{{\it Left}: The integrated \Lya\ SB maps within $\pm1500$\,\kms\ of the target redshift. From top to bottom, we present the brightest to faintest \Lya\ nebula respectively. {Top-Right}: Individual SB profiles for \idOne, \idTwo, \idThree, and \idFour, shown as blue, green, red, and orange squares, respectively. Error bars indicate the 25th$-$75th percentile range within each annulus. {\it Bottom-Right}: Mean \SQSO\ SB profile (black dots) compared with results from previous studies. The best-fit power law for the faint ($L_{\rm Bol}^{\rm QSO} = 45.28$) and the $L_{\rm Bol}^{\rm QSO}=46.34$ QSO sample from \citet{Gonzalez-Lobos25} are plotted as orange and blue lines respectively. The shaded regions correspond to their $1\sigma$ uncertainty. The profile of the SMG from \citetalias{Hall24} is shown as gold circles. }
    \label{fig:SB-Profiles}
\end{figure*}

To better compare the SMG-QSOs against each other we construct azimuthally averaged \Lya\ radial profiles. To construct these profiles, we first generate \Lya\ SB maps by integrating each datacube within a $\pm1500$\,\kms\ velocity window. Unlike the 3D moment maps, we do not mask pixels with S/N $<2$. These SB maps are shown in the left column of Figure\,\ref{fig:SB-Profiles} for each target. We then place logarithmically spaced circular annuli around each target to measure the average \Lya\ SB. The annuli are defined in angular distance units and converted to physical distance using the target redshift. With {\it CubeCarve}, we are able to probe the inner $\sim0-20$\,kpc without introducing strong artifacts, allowing us to use five annuli spanning from $R = [0,120]$\,pkpc. We account for cosmological SB dimming by scaling each profile by $((1+z_{\rm QSO})(1+2.6))^{4}$. The resulting radial profiles are shown in the bottom-left panels of Figure\,\ref{fig:SB-Profiles}.

A few notable differences emerge when comparing the four targets. \idOne\ and \idTwo\ in particular show contrasting behaviors: \idOne\ exhibits a relatively flat profile aside from a small enhancement near $10$\,kpc, while \idTwo\ displays a much steeper decline with radius. \idFour\ shows a similarly steep profile, whereas \idThree\ lies between these two extremes. These contrasts may reflect different physical environments, though direct comparison of individual profiles does not allow firm conclusions.

We now compare the SMG-QSO profiles with previously studied QSOs from the ongoing QSO MUSEUM. For this, the four \SQSO\ profiles are combined by averaging the SB in each annulus, and the uncertainties reflect the standard deviation within each bin. For the following radial bins: $[2.5, 11.3, 22.6, 45.3, 90.5 ]$ pkpc, we measure a mean \Lya\ SB of: $[8.6\pm3.9,\, 6.5\pm2.9,\, 4.0\pm1.6,\, 1.3\pm0.9,\, 0.5\pm0.4] \times 10^{-18}$\,\surf. In \cite{Gonzalez-Lobos25}, the authors divide their large QSO sample into several QSO luminosity bins, ranging from $45.28 < \log \left(L_{\rm Bol}^{\rm QSO}\right)< 47.53$. The mean QSO luminosity of our sample is $\log \left(L_{\rm Bol}^{\rm QSO}\right) = 46.6$\,erg\,s$^{-1}$, so we would expect to find the mean SMG-QSO profile to sit somewhere between the median brightness ($\log \left(L_{\rm Bol}^{\rm QSO}\right) = 46.34$\,erg\,s$^{-1}$) and the faintest ($\log \left(L_{\rm Bol}^{\rm QSO}\right) = 45.28$\,erg\,s$^{-1}$) QSO bin. Their best-fit exponential profiles are overlaid with the mean SMG-QSO profile in Figure\,\ref{fig:SB-Profiles}. 

Rather than finding the mean profiles in between the two extremes, they are below the faint QSO sample. This acts counter to the idea that the luminosity of the QSO is the primary driver of the brightness of the \Lya\ emission. The fainter profiles observed for the composite systems may be related to enhanced dust content in the host galaxies, which would suppress the available \Lya\ photon budget despite a clear line of sight for ionizing photons to escape into the CGM. However, if we look back at the individual profile for each target, we notice that only \idOne\ and \idThree\ are most distinct from the broader QSO sample. Meanwhile, \idTwo\ and \idFour\ resemble previous standalone QSOs. This suggests that the massive dust content is not the primary reason for the on average, fainter profiles. We require a larger sample of composite systems to better understand where they fit within the broader QSO sample.

\subsection{Host Galaxy Impacts on \Lya\ Emission}

In recent years, considerable effort has gone into measuring the cold molecular gas content of QSOs with known extended \Lya\ nebulae. As discussed previously, QSO host galaxies are known to occupy the most massive dark matter halos, with $M_{\rm DM}\sim 10^{12.5-13}$\,\msun\ based on clustering measurements \citep{DaAngela07,Hickox12,Trainor12,White12}. These same halos are also thought to host SMGs, pointing toward a possible evolutionary connection between the two populations. At redshifts $z\sim 2-3$, such halos are expected to sustain a mixture of hot ($10^{6-7}$ K) and cool gas phases in their CGM \citep{Keres05}. While the large-scale cool CGM is known to produce the extended \Lya\ emission seen around nearly all luminous QSOs \citep[e.g.][]{Arrigoni-Battaia19}, the role of the much colder ($10-100$ K) gas phase within QSO host galaxies remains less clear. Cold gas is the raw fuel for star formation, and understanding how it possibly connects to \Lya\ nebulae provides a potential link between the galaxy’s star-forming reservoir and its circumgalactic environment.

We begin our investigation into this possible link by searching for a connection between the far-IR emission and the total \Lya\ luminosity of the nebulae. Specifically, we examine the total integrated \Lya\ luminosity of each nebula as a function of the 870\,\um\ flux density ($S_{870}$ as reported in Table\,\ref{tab:lya_results}), where the far-IR emission traces the dust content and, by extension, the cold gas reservoir. We include the SMG from \citetalias{Hall24}, for which its $S_{870} = 7.4 \pm 0.5$\,mJy as measured by \citetalias{Fu17}. Because the KCWI data for this source reached a deeper surface brightness limit ($\sim 10^{-19}$\,\surf), we remeasured its \Lya\ luminosity by restricting the integration to regions with SB $> 1.6 \times 10^{-18}$\,\surf, matching the limits of the SMG-QSO observations. This yields $L_{\rm Ly\alpha} \approx 0.48 \times 10^{43}$\,erg\,s$^{-1}$. If we combine all of these properties together, we find that nebulae with brighter Ly$\alpha$ emission are associated with systems exhibiting lower $S_{870}$. A similar trend was reported by \citet{Gonzalez-Lobos23}, who examined the ratio of nebula to QSO \Lya\ luminosity $\left(L_{\rm Neb}^{\rm Ly\alpha} / L_{\rm QSO}^{\rm Ly\alpha}\right)$ as a function of dust mass inferred from $S_{870}$. In their study, the overall nebula integrated \Lya\ luminosity decreases with increasing dust mass, suggesting that dust plays a role in attenuating Ly$\alpha$ photons.

Yet, an equally plausible explanation is that the \Lya\ nebula properties trace a larger cycle that regulates the host galaxy’s cold gas reservoir. If these composites are indeed undergoing an evolutionary change from dusty starbursts to QSOs, then the CGM may not have the same cool gas content as seen in the faint and bright QSO sample. Specifically, there may simply be a lag between the QSO reaching its observed luminosity and material being expelled into the CGM. This ``blowout'' phase can deplete the host galaxy of its molecular gas required for star formation \citep{Menci08,Hopkins08}. This idea is supported through investigations into the cold gas content of QSOs \citep{Bischetti20}, revealing a lower molecular gas fraction than normal star forming galaxies. This suggests that AGN activation depletes this reservoir, which in turn could expel into the CGM. Recent works have begun searching for connections between the \Lya\ nebula and molecular gas content of $z > 2$ QSOs, such as the SUPERCOLD Survey \citep{Li23a} and the APEX at the QSOMUSEUM \citep{Munoz-Elgueta22}. Both of these studies have found conflicting results on the relationship between the integrated luminosity of the \Lya\ nebula and the total molecular gas mass of the host galaxy. Specifically, \cite{Li23a} finds that the total integrated luminosity of the \Lya\ nebulae increases as the cold gas content of the host galaxy increases. However, \cite{Munoz-Elgueta22} measures the opposite trend; the \Lya\ nebulae have lower luminosities as the cold gas mass increases. With the clear difference found in our \Lya\ SB profile and the apparent trend in $L_{\rm Ly\alpha}$ versus $S_{870}$, further investigations into the host galaxy of known \Lya\ nebula of QSOs is warranted.

\section{Summary and Conclusions}\label{sec:conclusions}

We analyzed four SMG-QSO systems using new KCWI observations and archival MUSE data. Each target represents a possible transitional phase between dusty starbursts and unobscured QSOs, offering insight into the physical processes that link these two populations. To enable a consistent comparison between the KCWI and MUSE datasets, we convolved and rebinned each MUSE datacube to match KCWI observations. The newly developed {\it CubeCarve} method, a modern update from \cite{Lucy94} focused on 3D data, provided a clean view of the extended \Lya\ emission within the CGM by iteratively modeling and removing the unresolved QSO component without direct subtraction or masking of the central regions.

Our main findings are as follows:

\begin{enumerate}
\item {\it CubeCarve} robustly extracts extended \Lya\ emission from both simulated and real IFU observations. In its current form, it can be applied to KCWI and MUSE data alike, allowing for consistent measurements of CGM emission without masking the bright QSO core.

\item \idOne\ hosts the faintest \Lya\ nebula in our sample, with a total integrated luminosity of $L_{\rm Ly\alpha} = (6.0\pm1.5)\times10^{42}$\,erg\,s$^{-1}$. Its azimuthally averaged \Lya\ surface brightness profile is also the shallowest, resembling the SMG studied in \citetalias{Hall24}. The low $L_{\rm bol}^{\rm QSO}$ may account for the faint nebular emission if the QSO dominates the ionizing budget.

\item \idTwo\ shows the brightest and broadest \Lya\ emission in the sample, with a total integrated luminosity of $L_{\rm Ly\alpha} = (2.5\pm0.13)\times10^{43}$\,erg\,s$^{-1}$ and a peak surface brightness of $\sim9\times10^{-18}$\,\surf. Its CGM exhibits distinct kinematic structure, with regions offset by $\pm400$\,\kms\ on opposite sides of the nebula, possibly tracing outflowing gas. The steep surface brightness profile resembles those of luminous QSO nebulae at similar redshifts.

\item \idThree\ displays a compact \Lya\ nebula with a peak surface brightness of $\sim4\times10^{-18}$\,\surf\ and $L_{\rm Ly\alpha}\approx1.4\times10^{43}$\,erg\,s$^{-1}$. No strong velocity gradients are detected, possibly due to the resampling procedure applied to the MUSE data.

\item \idFour\ exhibits a brighter and more extended nebula than \idThree, with an average surface brightness roughly twice as high and $L_{\rm Ly\alpha}\approx2.1\times10^{43}$\,erg\,s$^{-1}$. Most of the gas is redshifted relative to $z_{\rm QSO}$ by up to $\sim400$\,\kms.

\item Comparison between {\it CubeCarve} and traditional PSF-subtraction methods shows significant improvements, demonstrating the algorithm’s ability to recover faint emission without introducing artifacts.

\item On average, the SMG-QSOs display fainter and shallower \Lya\ surface brightness profiles than the general QSO population, closely matching the SMG in \citetalias{Hall24}.

\end{enumerate}

The development and application of {\it CubeCarve} mark a key step forward in analyzing faint, extended emission in IFU data. By systematically isolating spatially and spectrally coherent structures within a datacube, {\it CubeCarve} enables a more robust search for diffuse gas around complex systems such as SMG-QSOs. When applied to our sample, the technique reveals diverse morphologies and potential continuity between SMGs and QSOs.

The observed connection between far-IR luminosity and \Lya\ output suggests a complex interplay between the interstellar and circumgalactic media. These results underscore the importance of physically motivated, automated tools for uncovering low-surface-brightness structures that may otherwise remain hidden. Future applications of {\it CubeCarve} to larger IFU datasets and additional emission-line tracers will help build a more complete picture of how luminous QSOs shape their gaseous environments during the peak epoch of galaxy formation.

\section*{Acknowledgements}
This work is supported by the National Science Foundation (NSF) grants AST-1614326 and AST-2103251. Based on observations collected at the European Organisation for Astronomical Research in the Southern Hemisphere under ESO programmes 0103.A-0296(A), 0102.A-0403(A) and 0102.A-0428(A), available from the ESO Science Archive Facility https://archive.eso.org/. The authors wish to recognize and acknowledge the very significant cultural role and reverence that the summit of Maunakea has always had within the indigenous Hawaiian community. We are most fortunate to have the opportunity to conduct observations from this mountain.

{\it Facilities}: Keck/KCWI, KiDS, MUSE/VLT

{\it Software}: Astropy \citep{astropy22}, SciPy \citep{SciPy20}

\bibliography{smg-qso}
\bibliographystyle{apj}

\end{document}